# TOWARDS GREEN MEDIA DELIVERY: LOCATION-AWARE OPPORTUNITIES AND APPROACHES

HATEM ABOU-ZEID AND HOSSAM S. HASSANEIN, QUEEN'S UNIVERSITY

## ABSTRACT

Mobile media has undoubtedly become the predominant source of traffic in wireless networks. The result is not only congestion and poor Quality-of-Experience, but also an unprecedented energy drain at both the network and user devices. In order to sustain this continued growth, novel disruptive paradigms of media delivery are urgently needed.

We envision that two key contemporary advancements can be leveraged to develop greener media delivery platforms: 1) the proliferation of navigation hardware and software in mobile devices has created an era of location-awareness, where both the current and future user locations can be predicted; and 2) the rise of context-aware network architectures and self-organizing functionalities is enabling context signaling and in-network adaptation. With these developments in mind, this article investigates the opportunities of exploiting location-awareness to enable green end-to-end media delivery. In particular, we discuss and propose approaches for *location-based* adaptive video quality planning, in-network caching, content prefetching, and long-term radio resource management. To provide insights on the energy savings, we then present a cross-layer framework that jointly optimizes resource allocation and multi-user video quality using location predictions. Finally, we highlight some of the future research directions for location-aware media delivery in the conclusion.

## INTRODUCTION

The increasing popularity of online media content and video sharing through social networking websites is imposing a myriad of challenges to network operators. Mobile video is now forecasted to account for over 70 percent of the mobile data traffic by 2016. Much of this is pre-recorded video content such as movies, TV shows, and short clips delivered from popular sites such as YouTube and Netflix. As video content dominates the overall network traffic, it is also becoming a major contributor to the network energy consumption as well. Consequently, novel paradigms for green *end-to-end* video delivery are imperative to save energy across all the network elements.

Video delivery over 4G technologies such as Long Term Evolution (LTE) typically follows the architecture presented in Figure 1. When a video is accessed by a mobile device, it is first requested from content servers. The video stream then traverses the wireless carrier's core network (CN) and radio access network (RAN) before reaching the mobile user. Congestion at any point throughout the network results in video quality degradations, thereby reducing the perceived quality of experience (QoE). To cope with such challenges, several advancements have recently been proposed to the way media is delivered, including:

- adaptive video quality [1],
- in-network content caching [2],
- content prefetching into the user equipment (UE) local storage [3], and
- predictive radio resource allocations [4].

The goal of this article is to first review each of these key on-going developments, and then discuss opportunities and approaches of exploiting location-awareness at each network delivery level. This is motivated by recent analyses on human travel patterns indicating that people follow particular routes with a predictability of up to 88% [5]. These findings show that human mobility is highly dependent on historical behaviors. Moreover, for movement on buses and trains, mobility predictions can be facilitated even further.

In this article, we also shed light on the interactions between the delivery stages, and investigate how mobility patterns can be used to simultaneously coordinate efficient delivery. This is demonstrated through a proposal of *cross-layer* predictive green streaming, where radio resource allocation (RA) and video quality are jointly optimized using mobility predictions. Simulation results are provided and significant energy savings are observed by incorporating long-term location-awareness.

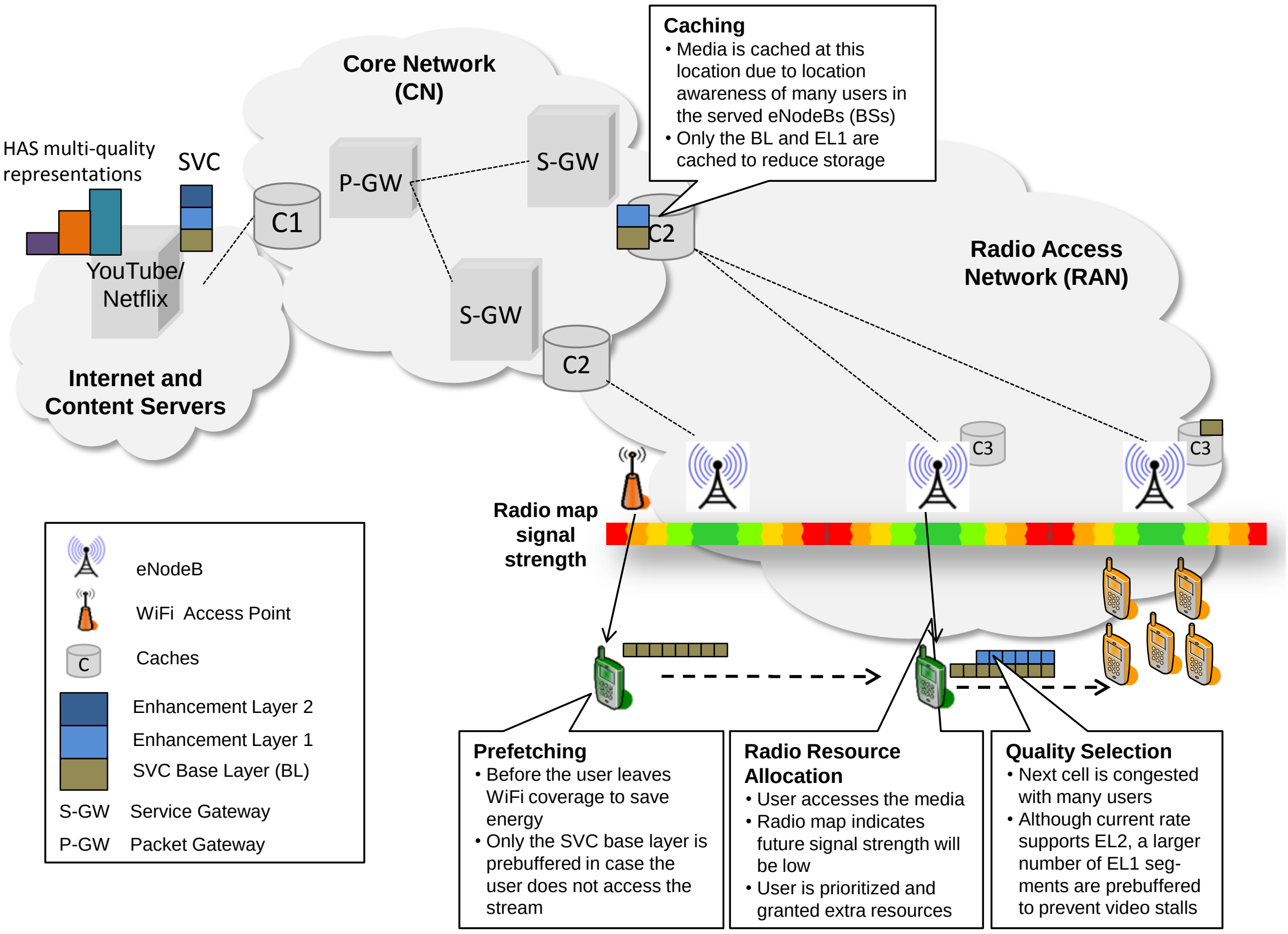

**Figure 1.** *Potential of location awareness and mobility predictions for end-to-end green media delivery.*

## The Era of Ubiquitous Location Awareness

LTE networks support a wide range of location positioning methods with varying levels of granularity, and a dedicated LTE Positioning Protocol (LPP) is also devised to coordinate signaling between the UE and the network. Concurrently, a plethora of navigation hardware and software is also available in vehicles and smart phones facilitating current location, as well as target destination reporting. Studies have also shown that the *prediction* of long-term user routes is possible by mining past trip observations [6], while information from road maps facilitates higher localization accuracy as well [7].

The derived user locations and trajectory predictions can be utilized in several respects. First, the spatio-temporal load distribution can be estimated in a network and hot-spots can be identified, thereby facilitating improved caching strategies. Additionally, if a user's future location is known, the upcoming data rates can be anticipated from radio maps that store average values of historic signal strengths at different geographical locations [8]. We envision that this era of ubiquitous location information can be leveraged to design efficient media delivery platforms as discussed below.

## Towards Green Media Delivery: Exploiting Location Awareness

In this section, we present key emerging developments in end-to-end media delivery over wireless networks. After discussing the pertinent challenges, we highlight how *location awareness* can be exploited to improve user experience and provide energy savings across the different network layers.

### Adapting Video Quality

Progressive download has become the most popular delivery mechanism for stored videos, accounting for over 50% of the Internet traffic in the US [1]. Media files are divided into fragments or chucks and most commonly delivered using HTTP over TCP. Once sufficient content is buffered at the receiver, the media file begins to play. In progressive streaming the media is typically delivered to the client at the same rate or 'quality level' irrespective of the fluctuating capacity of the core network and/or the wireless link. This results in video stalling when high quality segments are transmitted during bad channel conditions. To overcome this limitation, *multi-quality* video transmission has emerged:

***HTTP-based Adaptive Streaming* –** As in progressive streaming, media content is divided into a sequence of small file segments, each containing a short interval of playback time. The idea of HTTP-based Adaptive Streaming (HAS) is that each segment is then pre-encoded in multiple versions, each with a specific video bitrate and resolution. Higher quality segments will be larger in size but represent similar playback durations. The essence of HAS is that depending on the current network and wireless link capacity, the segment quality that can be timely received will be selected [1]. This reduces video stalls and is particularly suited for mobile streaming where users experience significant channel gain fluctuations. Several vendor specific variants of HAS have been developed in addition to the recent MPEG-DASH (Dynamic Adaptive Streaming over HTTP) standard [1].

***Scalable Video Coding* –** In Scalable Video Coding (SVC), media is encoded in multiple quality levels as in HAS. However, SVC uses a layered approach: a base layer (BL) is defined for the lowest video quality representation, and there are multiple enhancement layers (EL) above it. Each EL contains only the incremental data required to refine the video quality. The scalable layered approach permits the storage of one media file containing all the layers. This contrasts with having independent quality versions of the same media content as shown in Figure 1. The disadvantage however, is that SVC incurs additional processing overhead to combine the layers, introducing delays.

By utilizing SVC, a video can be decoded and viewed at the lowest quality if only the BL is delivered, while additional ELs improve the quality. SVC is resource efficient as a single representation containing the BL and ELs can be delivered from the server to the BS. Then depending on the link conditions, the appropriate number of layers is transmitted to each user. Several proposals have also been made to deploy SVC over DASH [9].

***Challenges* –** Typically, in HAS, each client tries to estimate the available bandwidth (by measuring the average arrival rate of data at the HTTP layer) and then choose the video rate accordingly. However, making accurate estimations can be challenging during congestion, resulting in poor user experience [1]. This is even more difficult in mobile environments due to the rapid link fluctuations and uneven spatial traffic distribution, causing frequent quality variations and video stalls. To overcome some of these challenges it is proposed to include centralized *in-network* adaptation frameworks where the BS determines the quality levels for each client and allocates the required radio resources [10].

***Energy Consumption* –** While quality adaptation was primarily developed to improve QoE [1], it influences energy consumption in several ways. Firstly, since smooth streaming reduces rebuffering delays, the device energy consumption will decrease due to less time spent receiving and displaying the content [11, 12]. Further, adaptive video streaming approaches allow the user to trade-off quality with energy savings if remaining battery power is low, even though the network conditions may permit high video quality. Similarly, the network may deliver lower quality videos to save content delivery energy.

***Location Awareness* –** Exploiting user mobility trajectories (and the associated rate predictions) to optimize HAS quality adaptation has been recently investigated with promising results. Yao *et al.* [13] develop an adaptation algorithm that proactively switches to the predicted transmission rates based on a stored bandwidth map. Similar maps are used to plan *long-term* segment qualities that ensure smooth streaming in [14]. With such an approach, a user headed to a tunnel will prebuffer several low quality segments in advance to prevent future video stalls. Figure 1 illustrates an additional use of location awareness. Here, although the UE *current* rate supports SVC-EL2, a larger number of EL1 segments are prebuffered to prevent future video stalling in the upcoming congested cell. Therefore, location predictions can provide valuable quality adaptation frameworks but more in-depth analysis is needed to evaluate their limitations.

## IN-NETWORK CACHING

Studies of 3G network traffic have revealed that popular content is generally requested by multiple users according to certain probability distributions [2]. Content caching enables the temporary storage of popular content and was typically performed close to mobile network operator gateways. However, recently, propositions for smaller caches located within the mobile network infrastructure have been made [2,15], as illustrated in Figure 1. Caching media closer to the mobile users reduces the need to re-deliver content from the original server, thereby decreasing the server and CN load, as well as the delivery delay [15].

| Efficient video delivery enabler | Description | Implementation challenges | Impact on energy | Directions for location-aware solutions |
|---|---|---|---|---|
| Adapting video quality | Seamlessly adapting video quality based on current network conditions to provide uninterrupted streaming. | Estimating end-to-end network conditions is difficult [1]. Mobility introduces more challenges from rapid fluctuations in wireless link capacities and spatial load densities. | Smooth streaming reduces rebuffering delays and device energy consumption. Lowering the target quality allows energy efficient delivery when energy is scarce [11,12]. | Making long-term segment quality plans based on: 1) future signal strength variations, and 2) spatial traffic distributions [14, 24]. |
| In-network caching | Storing popular media content at the edge of the network (i.e. closer to the user). This reduces the server and transport link load. | Predicting content popularity. Optimizing the size and location of caches in the network architecture to maximize the hit ratio and reduce required storage [2, 15]. | Reduces transport related energy consumption, but requires additional energy to power the caches [15]. | Current user locations and mobility trajectories can provide insight on where media is most likely to be accessed [15,17, 21]. |
| Prefetching/content pushing | Transmitting media content to the local storage of the UE before consumption, through less congested/more energy efficient delivery paths. | Determining the content to prefetch requires accurate user behavior modeling and mining [20]. Users may choose not to view the prefetched content, or view part of it [3]. | Prefetching over Wi-Fi and off-peak cellular network hours enables energy savings, but can incur energy waste if the content is not consumed [3,12,20]. | Knowing user locations can improve prefetching accuracy [17]. Mobility predictions also allow energy efficient scheduling of content pushing [4]. |
| Radio resource allocation | Assigning BS radio resources to minimize streaming interruptions. Information from user channel statistics and buffer queue length is used in RA algorithms. | Mobility and uneven traffic distribution causes sudden changes in the available data rates to users. Therefore, sustaining long-term QoE and fairness among users is difficult [1,23]. | Efficient RA schemes can reduce energy by sending media content in less time. RA strategies that prebuffer media to users allow the BS to enter sleep modes and save energy without impacting QoE [4,24]. | Rate predictions from mobility trajectories allow the BS to make efficient RA plans. Users are granted more resources when at peak channel conditions to opportunistically prebuffer content [23, 24]. |

**Table 1.** *Summary of key developments in end-to-end media delivery, and the directions for location-aware solutions.*

***Challenges –*** The efficiency of caching is generally determined by the cache hit ratio and the corresponding cache size. The objective is to maximize the media streaming experience while minimizing the additional costs incurred by cache servers. The challenge is to optimize the cache sizes and their geographical locations, as well as determine *which* content to store in the *different* network caches. The recent prevalence of media generated from SNs is anticipated to increase the importance of designing in-network caching strategies and architectures by facilitating better prediction of user media consumption [16, 17].

***Energy Consumption –*** By reducing the redundant data transfers between the content servers and the RAN, the effective core bandwidth of the network increases. Evidently, caching also reduces the delivery path and therefore the network related transmission energy [15]. At the end user, efficient reception (in less time) results in lower transceiver energy, and less rebuffering delays also reduce power consumption [11, 12]. On the other hand, the additional power consumption resulting from expanding the architectures of media caching will need to be evaluated to determine the trade-off.

***Location Awareness –*** Knowing the locations of users and their mobility patterns helps predict the network regions where media content will most likely be requested [17]. Figure 1 provides a use case of this where the appropriate CN cache is selected to store the media content. At a finer level, if the exact mobility trajectory of a user is known, it may be possible to determine the media segments to cache for that user in each BS depending on the predicted time period that will be spent at each cell [17]. We also foresee that mobility trajectories of users commuting on public transportation during rush hour can particularly be used to optimize the caching of popular evening shows and news reports at the appropriate locations.

It is worth noting that coupling location-aware caching with SVC based delivery can improve efficiency in many aspects. First, it is more storage efficient compared to having multiple quality representations as previously discussed. Secondly, hit ratios will be higher as it is very likely that at least the BL representation will be requested [18]. And finally, we foresee that it also enables hierarchical media caching [19], where for example only the BL can be cached at the BS, with additional ELs in the larger CN caches, as illustrated in Figure 1. This provides immediate access to the media stream (at a low quality), while enabling a gradual quality enhancement with acceptable delay.

## PREFETCHING/CONTENT PUSHING

In prefetching or content pushing, a part or the whole media content is loaded into the local storage of the UE before the user accesses the stream. This can be viewed as the extreme form of content caching that provides seamless streaming since the media is already preloaded [3, 12].

***Challenges –*** Prefetching is effective if the media content of interest is selected by the user in advance through subscriptions to playlists and other media services. However, the significant challenge is how to proactively determine the content users are anticipated to consume. As opposed to in-network caching which is based on aggregate access probability at a geographical location, prefetching requires insights on user-specific consumption preferences. This can be achieved with predictive approaches that monitor usage and apply data mining to identify usage patterns for news, specific shows, etc [20]. As in caching, information from SNs can be utilized in prefetching strategies where only the most popular content shared by the user's closest social circles are prefetching candidates [16]. Note that with SVC only the BL maybe prefetched in cases where the user access of the content is uncertain. This is illustrated in Figure 1.

***Energy Consumption –*** Prefetching media streams before consumption allows the content provider to choose less congested delivery paths (e.g. through WiFi) as well as off-peak hours to transmit the content to the users [12]. When content is transmitted in less time and/or through more energy efficient wireless interfaces, this translates to energy savings at the UE [11, 12] and the network. Furthermore, by distributing the traffic more evenly throughout the day, congestion from peak demands is reduced. On the other hand, prefetching incurs energy waste if the user does not access the pushed content [3].

***Location Awareness –*** As illustrated in Figure 1, being aware of a user's regular mobility patterns facilitates energy-efficient content pushing by proactively delivering content before a user leaves Femtocell or WiFi zones which consume less energy [12]. Similarly, predictions of upcoming high bandwidth or low congestion networks can delay content prefetching, and thereby optimize energy consumption [4]. With location awareness the spatio-temporal network load can also be distributed more evenly. Furthermore, coupling the users' content consumption profiles discussed early with their geographical locations may improve the accuracy of determining what to prefetch where [21]. However, more research is needed to propose effective, light-weight prefetching algorithms based on user profiling.

## RADIO RESOURCE ALLOCATION AND DISCONTINUOUS TRANSMISSION MODES

BSs transmit data to users by allocating units of bandwidth over time slots to different users. The goal of radio resource allocation (RA) is to distribute these bandwidth units efficiently and minimize video stalls or quality degradations. This is generally accomplished by allocating more resources to users with high channel conditions, while simultaneously ensuring that no users are starved.

***Challenges –*** Mobility poses a significant challenge in RA for video streaming. If a user remains in a poor radio coverage area or enters a congested zone, the achievable rates will not support the high video streaming requirements. Another important issue in adaptive streaming is that each client independently determines the segment quality level requested from the server. In traditional implementations, the BS is unaware of the selected quality levels and therefore does not know the target data-rates to transmit to each user. Additionally, since the clients make the quality selections independently, it may not be possible to satisfy them collectively. As previously mentioned, in-network video optimization where the BS determines the segment quality levels and makes the corresponding RAs have recently been proposed [10]. Maintaining long-term video quality fairness across users moving along multiple cells is another important challenge.

***Energy Consumption –*** BS power consumption is proportional to the supported traffic load [22]. This is because unused resources can be aggregated in a way that creates complete time slots without any data transmission, enabling transceiver hardware to be deactivated to save energy. Such mechanisms are known as the extended-cell discontinuous transmission (DTX) in Long Term Evolution (LTE) [22]. Similarly, the UE can also enter a discontinuous reception (DRX) mode to save power. Therefore, efficient media delivery RA schemes streaming will reduce energy consumption. Furthermore, a mode of deep sleep allows BSs to be dynamically switched off completely to save more energy, if no transmission is made for an extended period of time. In this sleep mode, a minimum off-period is needed before BSs can re-operate.

***Location Awareness –*** Radio maps discussed earlier enable predicting the data rates a user will experience if the mobility trajectory is known. With this rate information, the BS can plan spectrally efficient rate allocations without violating user streaming demands. For instance, if a user is moving towards the cell edge or a tunnel, the BS can increase the allocated wireless resources allowing the user to buffer more video segments. The idea is to grant users more air-time access at their highest achievable data rates and less access when they are at lower achievable rates. This allows the BS to transmit more data in less time, and consume lower transmit energy. Furthermore, a user on a highway traversing two cells, can be pre-allocated the requested video content in the first cell, while the second cell is switched off without causing any video stalling. More details on the potential and architecture of such a predictive access strategies are presented in [4], while [23, 24] discuss efficient predictive algorithms that minimize video degradations and ensure quality fairness among users.

In Table 1, we summarize the previous discussion on the vital role of location awareness as an enabler of green end-to-end media delivery. It is worth noting that to fully capitalize on location awareness, it is necessary to consider the interaction between the different layers of delivery and design cross-layer delivery strategies. To illustrate this and demonstrate the potential energy savings, we now present a case study on jointly optimizing RA and multi-user video quality using location predictions.

## LOCATION-AWARE CROSS-LAYER OPTIMIZATION OF RESOURCE ALLOCATION AND VIDEO QUALITY

In this section, we propose to use rate predictions (derived from mobility trajectories) to jointly plan RA and user video segment qualities over a time horizon. The objective is to minimize BS power consumption without causing any streaming stalls. We refer to this approach as predictive green streaming (PGS). Studies indicate that the major part of energy (50-80%) is consumed in the radio access [22]; therefore the savings from PGS will significantly reduce end-to-end delivery energy.

### SYSTEM MODEL

We consider a simple scenario of three cooperating BSs 1 km apart, covering a highway. Users request stored videos transmitted using HAS. Let $M$ denote the BS set $M=\{1,2,...,M\}$,and $N$ the active user set $N=\{1,2,...,N\}$. An arbitrary BS is denoted by $j$ and a user by $i$. Users are associated to BSs based on the closest distance. The set $U_{j,t}$ contains the indices of all the users associated with BS $j$ at time $t$. We generate vehicle traces using the SUMO microscopic traffic simulator[1]. Average user received power is computed based on the path loss model $PL(d)=128.1+37.6\log_{10}d$, where d is the user-BS distance in km[2]. The user-BS distance is assumed to be known during a *lookahead window* of $T$ seconds. Time is divided in equal slots of duration $\tau$, during which the path loss is assumed to be constant (we set $\tau = 1$ s). The link rate is computed using Shannon's equation with SNR clipping at 20 dB. We denote this rate by $r_{i,t}$ which is assumed to be known for the time window t=1,2,..$T$. This generates a matrix of *future* link rates defined by $\mathbf{r} = ( r_{i,t} : i \in N, t=\{1,2,...,T\})$. Figure 2(a) illustrates an example of $r_{i,t}$ for a user traversing two BSs along a highway. We assume that knowledge of $r_{i,t}$ is error free to provide the bounds of the potential gains, but further work is needed to investigate the effects of errors.

*1. Simulation of Urban Mobility (http://sumo.sourceforge.net)*
*2. Although we only consider path loss, a more complex channel model may be used as the proposed PGS approach is generic and only requires an estimate of the radio map which may even be determined either through measurements.*

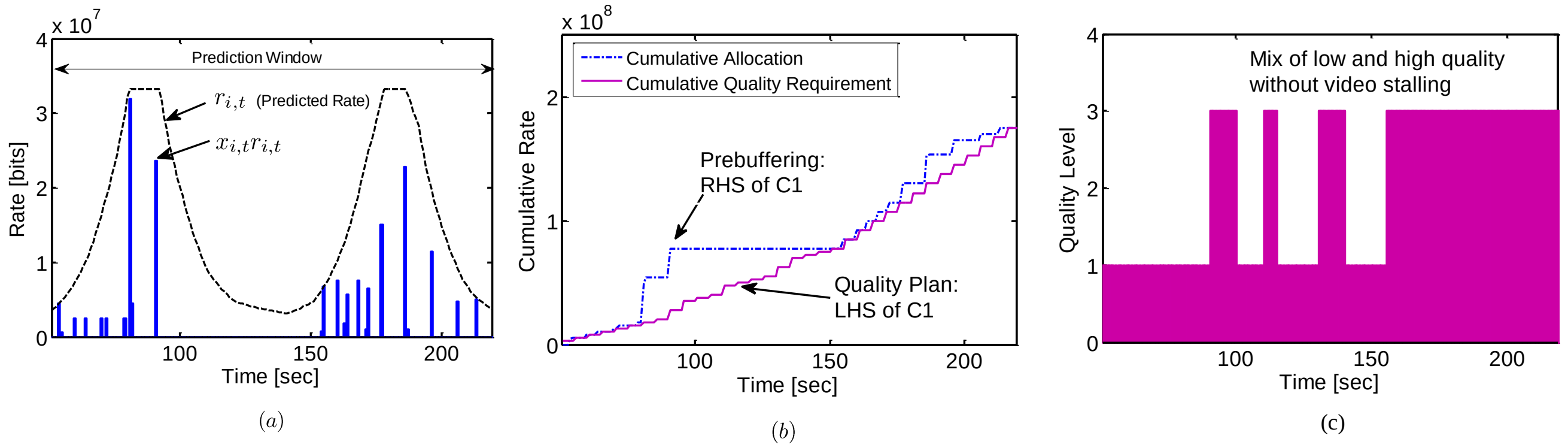

**Figure 2.** *Leveraging rate predictions to jointly optimize resource allocation and segment quality planning in adaptive media streaming.*

***Resource Sharing Factor –*** During $\tau$, BS air-time is shared in arbitrary ratios between the users. The resource sharing factor $x_{i,t} \in [0, 1]$ defines the fraction of time that the BS will serve user $i$ at time $t$. The received data rate is then $x_{i,t}r_{i,t}$.

***BS Power Model –*** BS downlink power is based on the linear load dependent model [22], where power is proportional to BS load, with a minimum power $P_0$ required at minimum load, and maximum power consumption $P_{\max}$ at maximum load.

***Adaptive Video Streaming Model –*** The duration of each video segment is $\tau_{seg}$ which is a multiple of $\tau$. The lookahead window $T$ is selected to be divisible by $\tau_{seg}$, and $S$ denotes the number of segments during $T$. Each segment is available in the quality levels defined by $Q=\{1,2,...,q_{\max}\}$. The function $f^Q(l)$ maps the quality level to the corresponding bitrate. To assign the quality level of each user segment, we define the binary decision array $\mathbf{q}=(q_{i,s,l}\in\{0, 1\}: i\in N, s=\{1,2,...,S\}, l \in Q)$, where $\sum_{l=1}^{q_{max}} q_{i,s,l} = 1$ to ensure that only one level is selected.

## Predictive Green Streaming: Problem Formulation

The proposed PGS leverages rate predictions to introduce the following novel features into adaptive media streaming:

***Efficient Resource Allocation –*** The idea is to allocate few resources to users that are far, but approaching the BS (just enough to sustain smooth streaming). Then, as channel conditions reach their peaks, significant video content is pre-buffered. By opportunistically determining when to pre-buffer excess content to each user, the transmit time can be reduced significantly, thereby saving BS and UE power. This idea is illustrated in Figure 2a where allocation is avoided during poor channel conditions. This allocation plan is made for all users in advance based on knowledge of the predicted rate matrix **r.**

***Long-term Quality Planning –*** In order to sustain smooth streaming, the video segment quality levels should not require more resources than that allocated by the BS. Therefore, when segments are prebuffered, the appropriate mix of high and low qualities should be selected such that the video buffer does not underflow before additional resources are granted. We illustrate this concept in Figure 2b and 2c. The joint relationship between the allocated rate and segment quality *plan* that ensures smooth playback can be captured in the following constraint:

$$\textbf{C1:} \quad \tau_{seg} \sum_{s'=1}^{s} \sum_{l=1}^{Q} q_{i,s',l}\, f^Q(l) \le \sum_{t=1}^{s\tau_{seg}} x_{i,t}\, r_{i,t} \quad \forall i, s$$

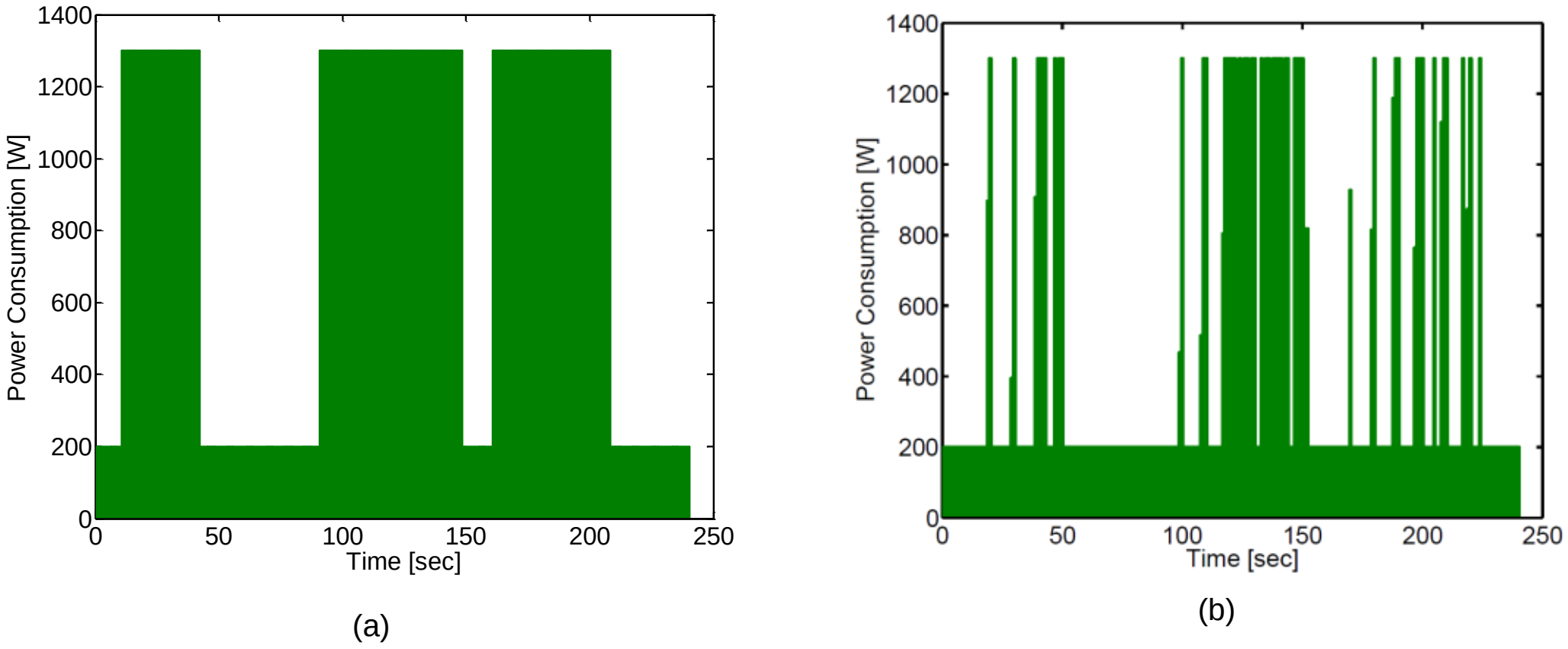

**Figure 3.** *Sample BS power consumption with time, where $P_{min}$=200 W and $P_{max}$=1300 W, and vehicular users arrive at the BS in three consecutive groups. (a) In traditional operation, BS airtime is fully utilized when users are present. (b) With PGS BS power is minimized by opportunistic allocations using rate predictions.*

where $\sum_{l=1}^{q_{max}} q_{i,s,l} = 1$. Note that we are interested in the *cumulative* allocations and qualities in order to account for pre-buffering. The right hand side of C1 gives the cumulative bits *allocated* to user *i* at the slots corresponding to the end of each segment, whereas the left hand side denotes the cumulative bits *required* to stream up to *s* video segments at the quality levels specified by $q_{i,s,l}$. For uninterrupted playback, segment *s* must be fully downloaded by time slot ‘$s\,\tau_{seg}$’. A limit can also be imposed on the number of pre-buffered bits, in case the user aborts the video early.

Constraint C1 ensures smooth playback but it does not set the target average quality level. To do so, we introduce the following constraint

$$\textbf{C2:} \quad \sum_{s=1}^{S} \sum_{l=1}^{q_{max}} q_{i,s,l} \geq l_{req} S \quad \forall\, i,$$

where $l_{req}$ denotes the desired average quality level for each user. Varying $l_{req}$ enables a trade-off between quality and energy consumption. Finally, we limit the sum of the air-time fractions allocated to the users to unity, i.e. **C3:** $\sum_{i \in U_{j,t}} x_{i,t} \leq 1 \quad \forall\, j, t$.

The optimal RA that minimizes BS power and achieves the target quality level without video stalling can therefore be derived by solving the following Mixed Integer Linear Program (MILP):

$$\begin{aligned} &\mathit{Minimize}: \sum_{i=1}^{N} \sum_{t=1}^{T} x_{i,t} \\ &\mathit{Subject\ to}: \quad \text{C1, C2, C3,} \\ &\qquad\qquad 0 \leq x_{i,t} \leq 1 \qquad \forall\, i, t. \\ &\qquad\qquad \mathrm{q}_{i,s,l} \in \{0,1\} \qquad \forall\, i, s, l. \end{aligned}$$

Note that this optimization problem, dubbed PGS, is coupled over multiple BSs. Thus, allocations in one BS, impact the future amounts allocated in upcoming BSs traversed by the user. Developing decentralized solutions is therefore an interesting direction to pursue in PGS, which we leave for future work. Furthermore, it is possible to make BS On/Off switching plans that turn the BS completely off at the times within *T* where no users require resources. With this approach, bulk user allocations can also be grouped together to prebuffer content and satisfy all the users before turning completely off as in [24].

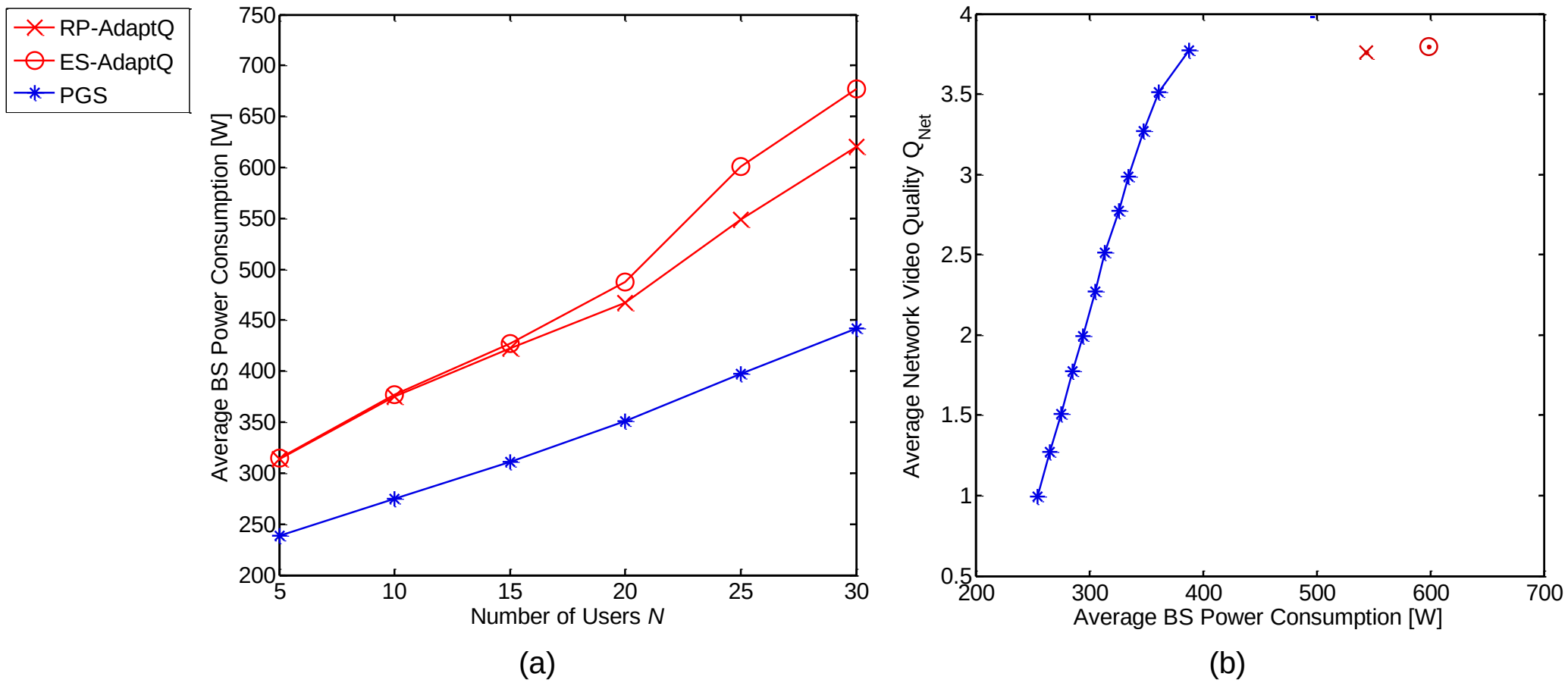

**Figure 4.** *(a) BS downlink power consumption for varying number of users; (b) the trade-off between average video quality and the BS power consumption.*

## Power Savings

We evaluate the potential power savings of PGS where the three BSs are 500m from the road edges and 1 km apart. Table 2 summarizes the network and video streaming parameters, and BS power consumption at minimum and maximum load is 200W and 1300W respectively, which is common for macro BSs employing time-domain duty cycling [22]. Gurobi 5.1[3] is used to solve PGS and Matlab was used as a simulation environment. Two baseline approaches that do not incorporate rate predictions are considered. The air-time allocation for these is based on:

***Equal Share (ES):*** air-time is shared equally among the users at each time slot.

***Rate-Proportional (RP)***: the RP allocator prioritizes users with high rates, while still serving users with poor rates. The user air-time is the ratio of its data-rate to the sum all users rates: $x_{i,t} = r_{i,t}/\sum_{i\in U_{j,t}} r_{i,t}$. Segment quality is then adapted based on the allocated rate at the start of the current segment, and the highest supportable level is selected. We refer to these approaches as ES-AdaptQ and RP-AdaptQ.

Figure 4a shows the average BS power consumption versus *N*. In this scenario all the allocators achieve a similar video quality $l_{req} \approx 3.75/4$ (which is not shown in the figure). PGS provides significant power savings, and at higher load it supports almost double the users compared to the baseline allocators when operating at the same power level. User mobility information allows the BSs plan efficient resource allocation by leveraging the rate prediction vector **r** as shown in Figure 2(b). This translates to energy savings as the BS is able to deliver the videos with less transmission time. In Figure 4(b) we present the additional power saving potential by reducing the target quality $l_{req}$ in the PGS optimization formulation. A lower video quality is represented with a lower bit-rate and therefore video segments transmission can be optimized further using rate predictions. The trade-off indicates that the power can be reduced by over 35% as the quality is decreased.

## Open Issues and Challenges

Research on location-aware media delivery is still in its infancy, and several open issues and challenges remain to be addressed.

***Predicting User Behavior:*** Using data analytics and user profiling to accurately predict what content users will consume and where is expected to revolutionize in-network caching and content pushing strategies. Efficient localization and mobility prediction algorithms are needed to facilitate the predictive approaches to content delivery.

***Modeling and Incorporating Uncertainty:*** After forecasting user location and content consumption, there is a need to investigate the effects of prediction errors on the developed location-aware schemes. Additionally, modeling prediction *uncertainty* and developing solutions that incorporate this uncertainty are required. Stochastic or fuzzy representations of the uncertainties are possible approaches. In such systems, feedback will be elemental to tuning the degree of uncertainty depending on the observed discrepancies from the predicted values. Estimation techniques and stochastic optimization may play an important role in designing robust predictive caching, prefetching and resource allocation.

[3] *Gurobi Optimization. http://www.gurobi.com/.*

| Parameter | Value |
|---|---|
| Bandwidth | 5 MHz |
| BS transmit power | 43 dBm |
| BS min power consumption | 200 W |
| BS max power consumption | 1300 W |
| Number of vehicles | 5 to 30 |
| Prediction window | 240 s |
| Vehicle arrival rate | 1 vehicle/s |
| Video quality bit-rates | {0.25,0.5,0.75,1} Mbit/s |
| Video segment duration | 10 s |

**Table 2.** *Simulation parameters.*

***Implementation Considerations:*** Studies on the practical integration into both current standards and emerging relevant developments such as network function virtualization and software defined networking are open research directions. As nodes become more programmable and scalable it would be interesting to see how location-awareness can dynamically configure/optimize network architecture and operation. Furthermore, large scale simulation studies with real road maps, vehicle trajectories, and traffic demand patterns will provide insight on more realistic performance measures and large-scale deployment and signaling issues.

## Conclusion

In this article, we highlighted the pivotal role of location-awareness in enabling greener end-to-end media delivery. Instead of focusing on the immediate network conditions, location awareness facilitates long-term resource planning and content delivery paradigms. Through this discussion we demonstrated how both QoE and energy can be significantly improved. We have also presented a study on joint RA and quality planning to emphasize the potential of cross-layer location-based delivery. However, since this is an emerging research topic, many open challenges and implementation issues need to be tackled in order to fully leverage the gains of location-awareness.

**Hatem Abou-zeid** is a postdoctoral fellow at the electrical and computer engineering department, Queen’s University. He received the Ph.D. degree from the same department, Queen’s University, Canada, in 2014. During hisprogram, he was awarded a DAAD RISE-Pro Research Scholarship in 2011 for a six-month internship at Bell Labs, Germany. Hatem is also an experienced Lecturer and has been granted several Teaching Fellowships at Queen’s University to instruct freshman and senior level engineering courses. His research interests include context-aware radio access networks, network adaptation and cross-layer optimization, adaptive video delivery, and vehicular communications. He is also a Technical Reviewer for several prestigious conferences and journals.

**Hossam S. Hassanein** is the Founder and Director of the Telecommunications Research Laboratory, School of Computing, Queen’s University, Canada, with extensive international academic and industrial collaborations. He is a leading authority in the areas of broadband, wireless and mobile networks architecture, protocols, control and performance evaluation. His record spans more than 500 publications in journals, conferences and book chapters, in addition to numerous keynotes and plenary talks in flagship venues. Dr. Hassanein served as the Chair for the IEEE Communication Society Technical Committee on Ad hoc and Sensor Networks (TC AHSN). He is a senior member of the IEEE, and an IEEE Communications Society Distinguished Speaker (Distinguished Lecturer 2008-2010). He has received several recognitions and Best Paper awards at top international conferences.